\documentclass[onecolumn,prl,aps]{revtex4}

\usepackage{amsbsy}
\usepackage{latexsym,epsfig,graphicx}
\usepackage{dcolumn}
\usepackage{graphicx}
\usepackage{subfigure}
\usepackage{comment}
\usepackage{color}
\usepackage{bm}
\usepackage{mathrsfs}
\usepackage{amsfonts}
\usepackage{amsmath}
\usepackage{color}
\usepackage{amssymb}
\usepackage{xspace}
\usepackage{epstopdf}
\graphicspath{{figures/}}
\usepackage{tabularx}
\usepackage{longtable}
\usepackage[normalem]{ulem}
\begin{document}
\title{Fast and robust production of quantum superposition states by the fractional shortcut to adiabaticity}
\author{Guan-Qiang Li\footnote{Email: liguanqiang@sust.edu.cn}, Hao Guo, Yu-Qi Zhang, Bo Yang, and Ping Peng}
\affiliation{Department of Physics and Institute of Theoretical Physics, Shaanxi University of
Science and Technology, Xi'an 710021, China}

\begin{abstract}
The fractional shortcut to adiabaticity (f-STA) for production of quantum superposition states is proposed firstly via a three-level system with $\Lambda$-type linkage pattern and a four-level system with tripod structure. The fast and robust production of the coherent superposition states is studied by comparing the populations for the f-STA and the fractional stimulated Raman adiabatic passage (f-STIRAP). The states with equal proportion can be produced by fixing the controllable parameters of the driving pulses at the final moment of the whole process. The effects of the pulse intensity and the time delay of the pulses on the production process are discussed by monitoring the populations on all of the quantum states. In particular, the spontaneous emission arising from the intermediate state is investigated by the quantum master equation. The result reveals that the f-STA exhibits superior advantages over the f-STIRAP in producing the superposition states.

\end{abstract}

\maketitle

\section*{I. Introduction}
Realizing fast and efficient quantum population transfer plays an important role in the fields of atomic physics, quantum optics, and quantum information~\cite{K.Bergmann1998, M.Shapiro2007, N.V.Vitanov2017, D.Guery-Odelin2019, THatomura2024}.  Many techniques for the quantum population transfer have been proposed in the past decades, where the stimulated Raman adiabatic passage (STIRAP) becomes one of the most popular techniques among them~\cite{U.Gaubatz1990, N.V.Vitanov1998, N.V.Vitanov2001, R.G.Unanyan2001}. The efficient realization of the quantum population transfer for three-level system from the initial state to the final one can be achieved by employing two partially overlapping delayed laser pulses with a counterintuitive sequence. High transfer efficiency requires that the adiabatic condition must be satisfied for the system during the whole process. From the view of experiments, the system needs a large pulse area for the driving pulses, which is equivalent to using a large pulse intensity or spending a very long transfer time~\cite{J.Klein2008, Y.-X. Du2014, K.S.Kumar2016}. Large pulse intensity in experiments may cause the damage of the experimental samples or equipments, and long transfer time duration may exceed the lifetime of the quantum system determined by the decoherence.

The fractional STIRAP (f-STIRAP) is introduced as a modification of the STIRAP technique, allowing for the transfer of a portion of the population from one state to the others in a three-level system. This results in population existing simultaneously in both of the states at least and formation of a superposed state~\cite{P.Marte1991, N.V.Vitanov1999, R.Unanyan1998, F.Vewinger2003, W.Huang2016, L.Yang2014, F.Dreisow2009-2}. The f-STIRAP has the same stability and sensitivity to the external environments
as the STIRAP. The population fraction can be achieved by controlling the intensity ratio of the driving pulses at final time. Ref. \cite{JChath2023} shows that the adiabatic creation of the maximally coherent superposition is possible even in the absence of the two-photon resonance by chirping the pump and Stokes pulses equally and introducing a chirping delay in the second Stokes pulse for a four-level system. The mechanism has become a powerful tool for producing the superposition states and has been applied to the systems such as the superconductor quibits~\cite{GSParaoanu2019,Y.Yu2021}, ultracold atoms~\cite{I.Stevenson2023,CanmingHe2023}, acoustic metamaterials~\cite{S.Tang2022-2} and optical waveguides~\cite{M.Amniat-Talab2005,F.Dreisow2009,S.Fan2023}. But, the f-STIRAP still has the similar disadvantages with the STIRAP due to the requirement for the adiabaticity.

For revolting the adiabatic requirement, several strategies for accelerating the STIRAP have been proposed, including the transitionless quantum driving~\cite{M.Demirplak2003, M.Demirplak2008, M.V.Berry2009}, fast-forward scaling~\cite{S.Masuda2008, E.Torrontegui2012, K. Takahashi2014, JJZhu2021}, inverse engineering based on Lewis-Riesenfeld invariants~\cite{H.R.Lewis1969, X.Chen2011}, variational method~\cite{X.Chen2012,D.Sels2017}, and so on. Some of them are interrelated and can potentially be made equivalent by appropriately adjusting the reference Hamiltonian. Based on their common characteristics, these methods and techniques were also referred to as shortcut to adiabaticity (STA)~\cite{D.Guery-Odelin2019, THatomura2024, XiChen2010, E.Torrontegui2013}. Introducing time-dependent auxiliary pulses to eliminate the system's nonadiabatic couplings, the STA can realize alternative fast processes with the same population of the final state or even the same final state as the STIRAP in a finite short time~\cite{Y.X.Du2016}. It does't require that the system must evolve along an adiabatic pathway and has better stability for the parameters of the system, at the cost of applying predesigned auxiliary pulses matched with the transfer process. The applications of the STA have been extended to many classical and quantum systems, such as acoustic waveguides~\cite{S.Tang2022-1}, optical waveguides and lattices~\cite{R.Alrifai2023,V.Evangelakos2023, X.Liu2024,D.Stefanatos2014,G.D.Valle2018}, trapped ions~\cite{E.Torrontegui2018, J.Cohn2018}, quantum dots~\cite{XFLiu2024}, superconducting circuits~\cite{WZheng2022,F.Setiawan2023}, and so on. Recently, the related theoretical studies have also been generalized to multilevel systems, including the chainwise and $N$-pod systems~\cite{M.Amniat-Talab2011,M.Saadati-Niari2016,M.Saadati-Niari2020,N.Irani2022, R.Vahidi-Asl2023}.

 Similar to the relationship between the STA and the STIRAP,  the fractional STA (f-SAT) can overcome the shortcomings existing in the f-STIRAP, which needs high requirement for the adiabatic conditions. The transfer time is shorter than the decoherence time of the system for the f-STA. Despite its superiority, the concept of the f-SAT has never been put forward as far as we know. In this paper, the f-STA is proposed firstly via a three-level system with $\Lambda$-type linkage pattern and a four-level system with tripod structure and its effectiveness and reliability for the production of the superposition states is verified. The f-STA can transfer the quantum population from one ground state to the others and realize coherent superposition with arbitrary proportion quickly by adjusting the ratio of the coherent pulse intensities. It is shown that the f-STA can bypass the adiabatic requirements, and is more robust than the f-STIRAP by analyzing the effects of the pulse intensity, the time delay and the spontaneous emission on the transfer process.

The present paper is organized as follows. In Sec.~II, the f-STA of the three-level system for coherently producing the quantum superposition state with the population ratio of $\frac{1}{2}:\frac{1}{2}$ is proposed, and the effects of the pulse intensity, the time delay and the spontaneous emission on the population transfer are investigated. In Sec.~III, the f-STA of the four-level system is implemented for achieving superposed state with a population ratio of $\frac{1}{3}:\frac{1}{3}:\frac{1}{3}$, and the stability of the f-STA under the influence of the pulse intensity, the time delay and the spontaneous emission is discussed. Due to the adjustability of the parameters, the superposition states of any population ratio can be produced by the f-STA in the three- and four-level systems. The multilevel systems, such as N-pod linkage ($N\geq4$), may be reduced into the effective three-level systems by the Morris-Shore transformation~\cite{MSTransformation2023,BWShore2014,ESKyoseva2006,KNZlatanov2020}. The mechanism can be generalized to produce the superposition states with more than three states. The conclusion is given in Sec.~IV.

\section*{II. Fractional STA for the three-level system}
\subsection{A. Theoretical framework}
\begin{figure}[h]
\centering
\includegraphics[width=0.55\textwidth]{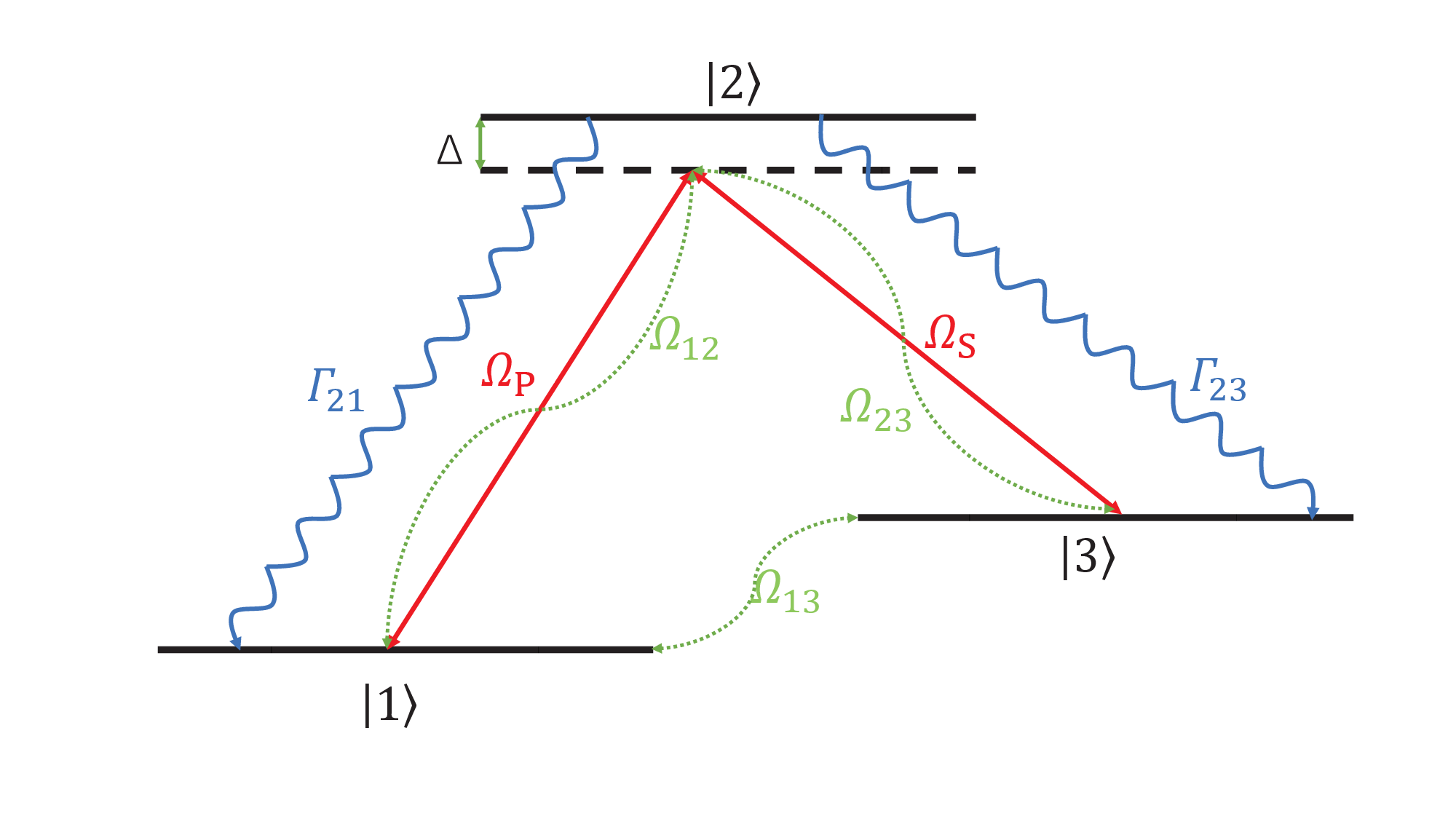}
\caption{Scheme of the population transfer for the three-level system. The coupling scheme is driven by the pump pulse $\Omega_{P}$ and Stokes pulse $\Omega_{S}$ (Red solid lines) for both of the f-STA and the f-STIRAP. The auxiliary pulses $\Omega_{12}$, $\Omega_{23}$ and $\Omega_{13}$ depicted by the dotted green lines are designed only for the f-STA. $\Gamma_{21}$ and $\Gamma_{23}$ represent the spontaneous emissions from the intermediate state $|2\rangle$ to the states $|1\rangle$ and $|3\rangle$ (Blue wavy lines). }
\label{figure1}
\end{figure}
Let's first consider the scheme of the f-STIRAP and the f-STA by a three-level system with $\Lambda$-type linkage pattern as shown in Fig.~\ref{figure1}. $|1\rangle$ and $|3\rangle$ correspond to the ground and metastable states, while $|2\rangle$ represents the intermediate excited state. The pump pulse $\Omega_{P}(t)$ and the Stokes pulse $\Omega_{S}(t)$ relate the states $|1\rangle$ and $|3\rangle$ to the excited state $|2\rangle$. In the rotating-wave approximation, the Hamiltonian of the system reads~\cite{U.Gaubatz1990, N.V.Vitanov1998, N.V.Vitanov2001,R.G.Unanyan2001}
\begin{equation}
\widehat{H}_{0}(t)=\frac{\hbar}{2}\left(
\begin{array}{lll}
~~~0&~~~\Omega_{P}(t)&~~~~~0\\
\Omega_{P}(t)&~~~\Delta(t)&~~\Omega_{S}(t)\\
~~~0&~~~\Omega_{S}(t)&~~~~~0\\
\end{array}
\right).\label{Hamiltonian.threeL.1}
\end{equation}
The states $|1\rangle$ and $|3\rangle$ are on two-photon resonance and the state $|2\rangle$ is off-resonance from the others by a certain detuning $\Delta(t)$. The wavefunction $|\psi(t)\rangle\equiv c_{1}(t)|1\rangle+c_{2}(t)|2\rangle+c_{3}(t)|3\rangle$ follows the Schr$\ddot{o}$dinger  equation:
\begin{equation}
\mathrm{i} \hbar \frac{\mathrm{d}}{\mathrm{d} t}|\psi(t)\rangle=\widehat{H}_{0}(t)|\psi(t)\rangle. \label{schrodingereq.1}
\end{equation}
Theoretically, the system is only populated on the state $|1\rangle$ at the initial time, i.e., $c_{1}(-\infty)=1$, $c_{2}(-\infty)=0$ and $c_{3}(-\infty)=0$, and the dynamics of the system is investigated by monitoring the populations $|c_{n}(+\infty)|^{2}$ $(n=1, 2, 3)$ on all of the states at $t\rightarrow+\infty$.

There are three eigenvalues for the Hamiltonian~(\ref{Hamiltonian.threeL.1}):
 $\lambda_{0}(t)=0$ and $\lambda_{\pm}(t)=\hbar[\Delta(t)\pm\sqrt{\Delta^{2}(t)+\Omega^{2}(t)}]/2$ with  $\Omega(t)=\sqrt{\Omega_{P}^{2}(t)+\Omega_{S}^{2}(t)}$.  The dressed eigenstate corresponding to $\lambda_{0}(t)$ is called the dark state: $|\lambda_{0}(t)\rangle=[\cos\theta(t),0,-\sin\theta(t)]^{T}$. The dressed eigenstates corresponding to the remaining eigenvalues are called the bright states: $|\lambda_{+}(t)\rangle=[\sin\phi(t)\sin\theta(t),\cos\phi(t),\sin\phi(t)\cos\theta(t)]^{T}$ and
 $|\lambda_{-}(t)\rangle=[\cos\phi(t)\sin\theta(t),-\sin\phi(t),\cos\phi(t)\cos\theta(t)]^{T}$ with the mixing angles $\theta(t)$ and $\phi(t)$ defined by $\tan\theta(t)=\Omega_{P}(t)/\Omega_{S}(t)$ and $\tan2\phi(t)=\Omega(t)/\Delta(t)$.
The dark state, which is a linear superposition of the pure states $|1\rangle$ and $|3\rangle$, is used as a passage to produce the coherent superposition state. The proportions are $\cos^{2}\theta$ in the state $\left | 1 \right \rangle$ and $\sin^{2}\theta$ in the state $\left|3\right\rangle$ for the whole transfer process. From the representation of the dressed states, the transfer process based on the dark state can be effectively implemented by employing two partially overlapping delayed Gaussian pulses $\Omega_{P}(t)$ and $\Omega_{S}(t)$ with a counterintuitive sequence. The counter intuition means the Stokes pulse $\Omega_{S}(t)$ is opened before the pump pulse $\Omega_{P}(t)$. This is the mechanism of the STIRAP~\cite{K.Bergmann1998, M.Shapiro2007, N.V.Vitanov2017}. For the complete transfer of the populations from $\left | 1 \right \rangle$ to $\left|3\right\rangle$, the adiabatic condition must be satisfied, which will spend very long transfer time.

The f-STIRAP can be obtained from the incomplete population transfer of the STIRAP. The variation of the mixing angle during the transfer process is controlled by appropriately designing the pulse sequence. Unlike the STIRAP, here the two pulses vanish simultaneously while maintaining a fixed ratio~\cite{N.V.Vitanov1999}:
\begin{equation}
\lim _{t \rightarrow-\infty} \frac{\Omega_{p}(t)}{\Omega_{S}(t)}=0, \quad \lim _{t \rightarrow+\infty} \frac{\Omega_{p}(t)}{\Omega_{S}(t)}=\tan \alpha. \label{auxiliarypulse.3}
\end{equation}
The final proportion of the superposition state is determined by the controllable parameter $\alpha$ in the driving pulses. Associated with the definition of the mixing angle $\theta(t)$, $\theta(t\rightarrow+\infty)=\alpha$ must be obtained. The specific form of the pulses for the f-STIRAP can be designed as follows~\cite{N.V.Vitanov1999, L.Yang2014, W.Huang2016}:
\begin{equation}
\begin{array}{c}\Omega_{P}(t)=\Omega_{0} \sin \alpha e^{-(t-\tau)^{2} / T^{2}},
\\
\\
\Omega_{S}(t)=\Omega_{0} [e^{-(t+\tau)^{2} / T^{2}}+\cos\alpha e^{-(t-\tau)^{2} / T^{2}}].\end{array}\label{auxiliarypulse.4}
\end{equation}
$\Omega _{0}$ represents the peak pulse intensity, $\tau$ represnts the time delay between the pump and the Stokes pulses, and $T$ represents the common duration of the pulses.

The f-STIRAP encounters the drawbacks of requiring a large peak pulse intensity $\Omega_{0}$ and/or longer duration due to the requirement of the adiabatic conditions. In general, it's hard to be realized in realistic platforms. To address this issue, the f-STA is adopted, which introduces the auxiliary pulses to couple the eigenstates, forming a closed loop and eliminating the influence of the nonadiabatic transitions. For obtaining the auxiliary pulses in the STA, the wavefunction $|\tilde{\psi}(t)\rangle=\widehat U(t)|\psi(t)\rangle$ follows the time-dependent Schr$\ddot{o}$dinger equation:
\begin{equation}
i\hbar \frac{\partial |\tilde{\psi}(t)\rangle }{\partial t} =\left [\widehat{U}^{\dagger}(t)\widehat{H}(t)\widehat{U}(t)-i\hbar \widehat{U}^{\dagger}(t)\frac{\partial \widehat{U}(t)}{\partial t}\right ]|\tilde{\psi}(t)\rangle.
\label{MotionEq.1}
\end{equation}
The second term on the right-hand side of the above equation,  i.e., $i\hbar \widehat{U}^{\dagger}(t) \frac{\partial \widehat{U}(t)}{\partial t}$, represents the influence of the nonadiabatic transitions. In order to speed up the production of the superposition states efficiently and robustly, it is necessary to properly induce an auxiliary Hamiltonian $\widehat{H}_{a}(t)$, which ensures that only diagonal term $\widehat{U}^{\dagger}(t)\widehat{H}(t)\widehat{U}(t)$ exist in the right-hand side of Eq.~(\ref{MotionEq.1})~\cite{D.Guery-Odelin2019, THatomura2024, XiChen2010, E.Torrontegui2013}. In other words, we need to make $ \widehat{U}^{\dagger}(t)\widehat{H}_{a}(t)\widehat{U}(t)-i\hbar \widehat{U}^{\dagger}(t)\frac{\partial \widehat{U}(t)}{\partial t}=0$, which leads to:
\begin{equation}
\widehat{H}_{a}(t)=i \hbar \frac{\partial \widehat{U}(t)}{\partial t} \widehat{U}^{\dagger}(t). \label{auxiliarypulse}
\end{equation}

Using the transformation matrix
\begin{equation}
\widehat{U}(t)=\left(\begin{array}{ccc}
\sin\phi(t)\sin\theta(t) & \cos\theta(t) & \cos\phi(t)\sin\theta(t) \\
\cos\phi(t) & 0 & -\sin\phi(t) \\
\sin\phi(t)\cos\theta(t) & -\sin\theta(t) & \cos\phi(t)\cos\theta(t)
\end{array}\right),
\label{transformation matrix.1}
\end{equation}
the auxiliary Hamiltonian in the three-level system can be derived as
\begin{equation}
\widehat{H}_{a}(t)=i\left(\begin{array}{ccc}0 & \Omega_{12}(t) & \Omega_{13}(t) \\ -\Omega_{12}(t) & 0 & \Omega_{23}(t) \\ -\Omega_{13}(t) & -\Omega_{23}(t) & 0\end{array}\right), \label{auxiliarypulse.1}
\end{equation}
with the auxiliary pulses
\begin{equation}
\Omega_{12}(t)=\dot{\phi}(t)\sin\theta(t), ~~~\Omega_{23}(t)=-\dot{\phi}(t)\cos\theta(t), ~~~\Omega_{13}(t)=\dot{\theta}(t). \label{auxiliarypulse.2}
\end{equation}
Here, $\dot{\theta}=(\dot{\Omega}_{P}\Omega_{S}-\dot{\Omega}_{S} \Omega_{P})/\Omega^{2}$ and $\dot{\phi}=\Delta(\dot{\Omega}_{P} \Omega_{P}+\dot{\Omega}_{S}\Omega_{S}) /[\Omega(\Delta^{2}+4 \Omega^{2})]$. The Hamiltonian $\widehat{H}_{0}(t)$ in Eq.~(\ref{schrodingereq.1}) is substituted by $\widehat{H}(t)=\widehat{H}_{0}(t)+\widehat{H}_{a}(t)$ for implementing the STA. The introduction of the auxiliary pulses allow for the elimination of the nonadiabatic transitions, providing a shortcut to bypass the adiabatic condition~\cite{D.Guery-Odelin2019,THatomura2024}. It also avoids the need for large pulse intensities while the production of the superposition states is still robust and efficient. The f-STA originates from the combination of the f-STIRAP and the STA.

\subsection{B. Coherent production of the superposition state }
The population transfer in the f-STIRAP is obtained by directly applying the pulse sequence~(\ref{auxiliarypulse.4}) to Eqs.~(\ref{Hamiltonian.threeL.1}) and (\ref{schrodingereq.1}). The transfer for the f-STA is realized by solving Eq.~(\ref{schrodingereq.1}) but the Hamiltonian $\widehat{H}_{0}(t)$ is substituted by $\widehat{H}(t)=\widehat{H}_{0}(t)+\widehat{H}_{a}(t)$. The auxiliary Hamiltonian $\widehat{H}_{a}(t)$ is provided by Eqs.~(\ref{auxiliarypulse.1}) and (\ref{auxiliarypulse.2}). Producing the superposition states with arbitrary proportion are possible but we first examine the most interesting case of equal proportion for the states $|1\rangle$ and $|3\rangle$. By fixing the parameter $\alpha=\pi/4$, the final state can be obtained as $|\psi(+\infty)\rangle=(|1\rangle-|3\rangle)/\sqrt{2}$, which corresponds to the population ratio of $\frac{1}{2}:\frac{1}{2}$ between $|1\rangle$ and $|3\rangle$. Such state will be used for example in building Raman quibit gates for quantum computation~\cite{BTTorosov2020,SGStanchev2024}.
\begin{figure}[h]
\centering
\includegraphics[width=0.33\textwidth]{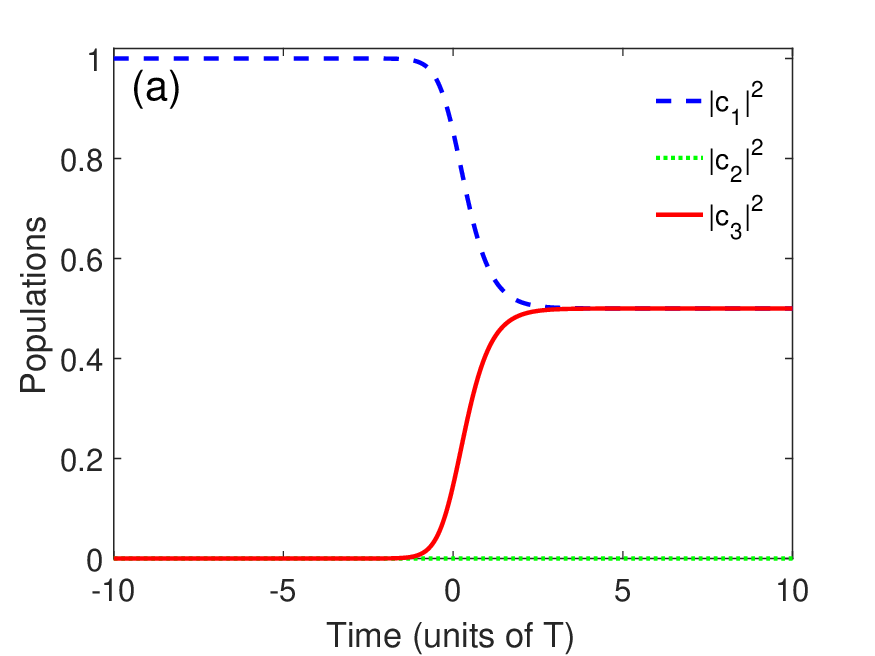}
\includegraphics[width=0.33\textwidth]{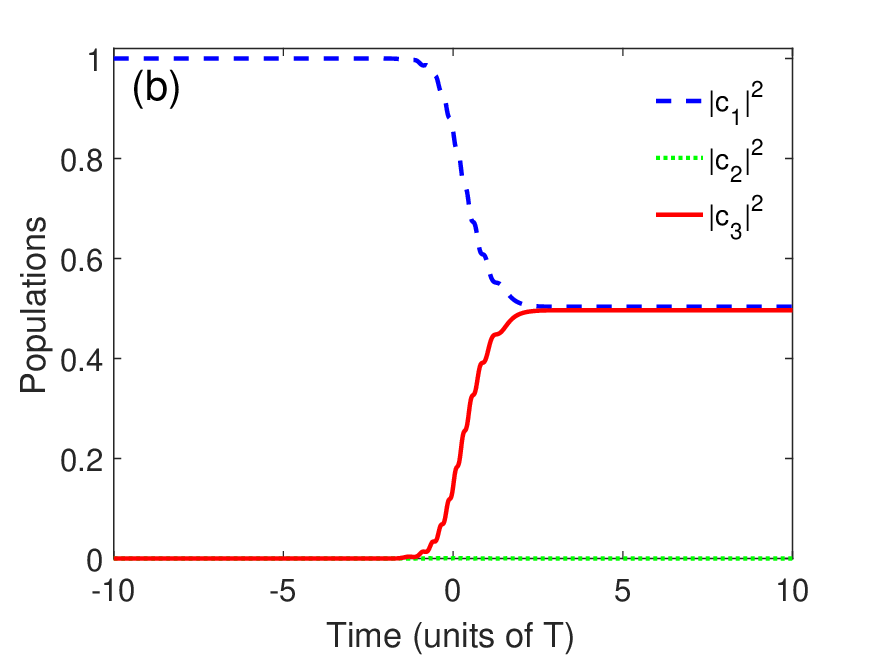}
\includegraphics[width=0.33\textwidth]{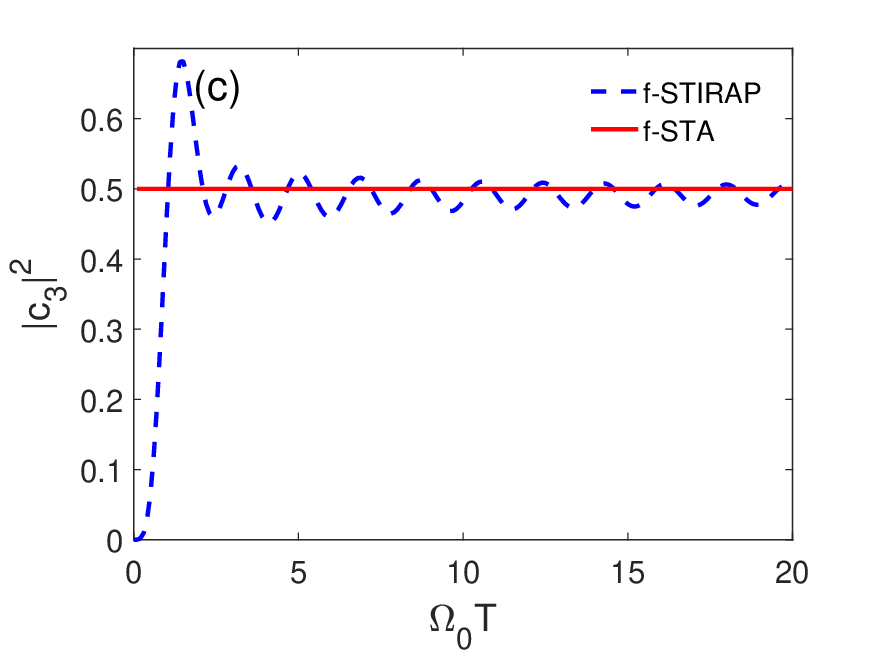}
\includegraphics[width=0.33\textwidth]{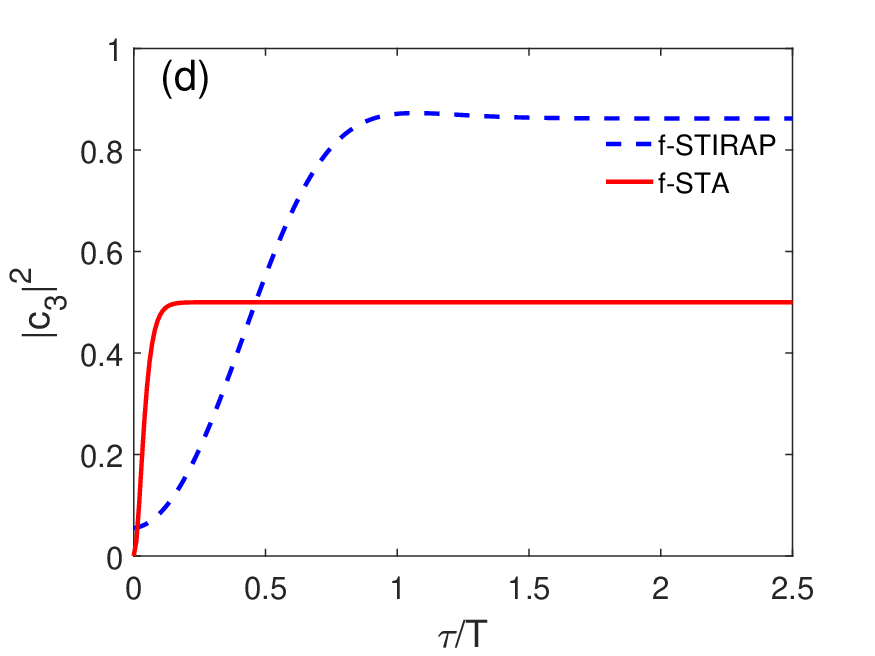}
\caption{Population transfer in the three-level system via the dark-state passage for the f-STA in (a) and for the f-STIRAP in (b). The parameters are chosen as $\alpha=\pi/4$, $\Omega_0T=2.0$, $\tau/T=0.7$ and $\Delta=0.2\pi$. The effects of the peak pulse intensity $\Omega_{0}$ in (c) at $\tau=0.7T$ and the time delay $\tau$ in (d) at $\Omega_0T=2.0$ on the final population on $|3\rangle$ for both of the f-STA (Red solid line) and the f-STIRAP (Blue dashed line).  }\label{figure2}
\end{figure}

The evolution of the populations for the f-STA and the f-STIRAP via the dark-state passage $|\lambda_{0}(t)\rangle$ is demonstrated in Fig.~\ref{figure2}(a) and (b), respectively. The population is assumed to stay on the state $|1\rangle$ at the initial time, and then $50\%$ of it is transferred  to the state $|3\rangle$ at the final time for the f-STA in Fig.~\ref{figure2}(a). There is no population on the state $|2\rangle$ during the whole process. By contrast, the final population on $|3\rangle$ doesn't reach $0.5$ for the f-STIRAP in Fig.~\ref{figure2}(b), because there is some population lost due to the nonadiabatic transitions. The f-STA introduces three auxiliary pulses (\ref{auxiliarypulse.2}) to eliminate the nonadiabatic transitions, perfectly achieving a coherent superposition state of $|1\rangle$ and $|3\rangle$ with equal proportion.

Fig.~\ref{figure2}(c) shows the effect of the peak pulse intensity $\Omega_{0}$ on the population transfer. The final population on $|3\rangle$ for the f-STA doesn't change with $\Omega_{0}$ and always keeps at $0.5$ for the f-STA. The transfer is not sensitive to the change of the pulse intensity. However, the final population on $|3\rangle$ for the f-STIRAP increases rapidly with increasing the pulse intensity first, then drops to $0.5$ after reaching a peak, and finally oscillates around $0.5$ with characteristic of amplitude attenuation. Obviously, the transfer for the f-STIRAP is affected dramatically by the change of the pulse intensity. Fig.~\ref{figure2}(d) shows the effect of the time delay $\tau$ on the population transfer. The final population on $|3\rangle$ for the f-STA increases rapidly with increasing $\tau$ and reaches to $0.5$ during a small time delay. For the larger time delay, the population always keeps at $0.5$. It means the presence of the three auxiliary pulses lessens the requirement that the two driving pulses must overlap. The final population on $|3\rangle$ for the f-STIRAP increases with increasing the value of $\tau$, but the growth of the population is relatively slow, and $50\%$ of the population can be achieved only at $\tau\approx0.47T$. With increasing the time delay, the population on $|3\rangle$ continues to rise and stabilizes after reaching to about $0.86$. As the time delay is increased, the overlap between the two driving pulses becomes smaller and smaller. Eventually, the two pulses no longer overlap, leading to a situation where the adiabatic condition is not satisfied. Consequently, the f-STIRAP can't give reliable results for larger time delay.

\begin{figure}[h]
\centering
\includegraphics[width=0.45\textwidth]{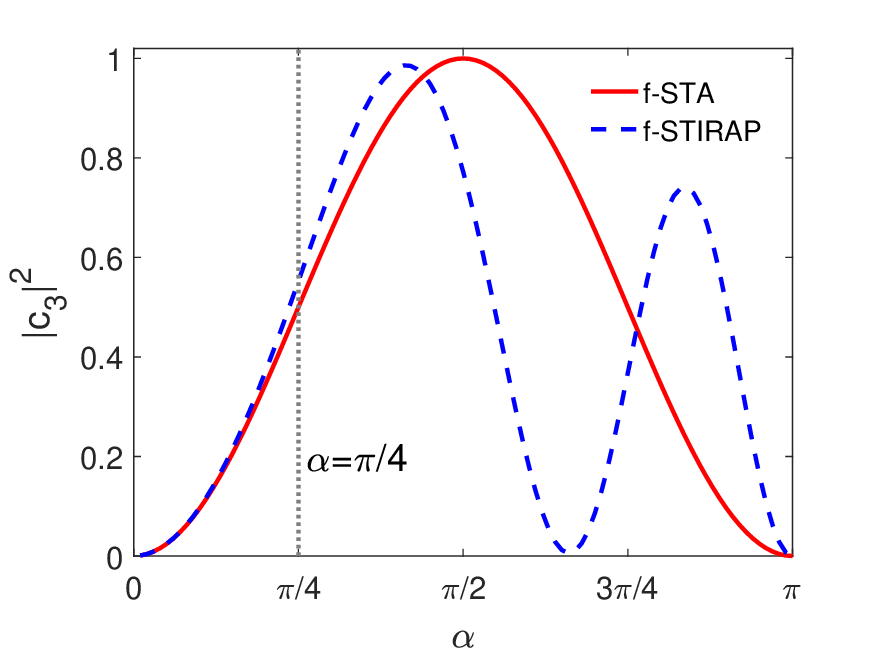}
\caption{ Change of the final population on $|3\rangle$ with the parameter $\alpha$ for the f-STA (Red solid line) and the f-STIRAP (Blue dashed line). The vertical dotted line corresponds to $\alpha=\pi/4$. The other parameters are chosen as $\Omega_0T=2.0$, $\tau/T=0.7$ and $\Delta=0.2\pi$.  }
\label{figure3}
\end{figure}

\begin{figure}[h]
\centering
\includegraphics[width=0.4\textwidth]{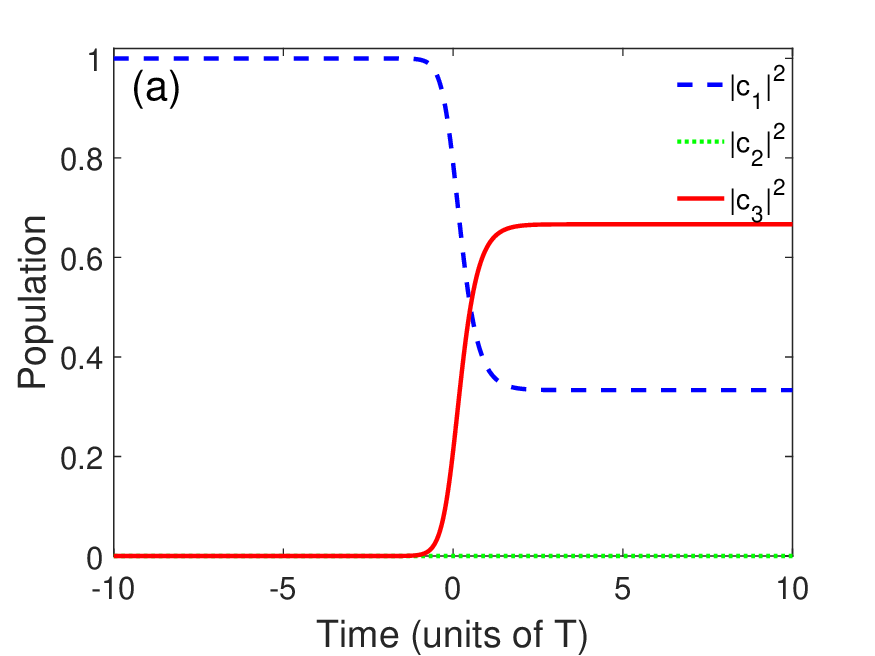}
\includegraphics[width=0.4\textwidth]{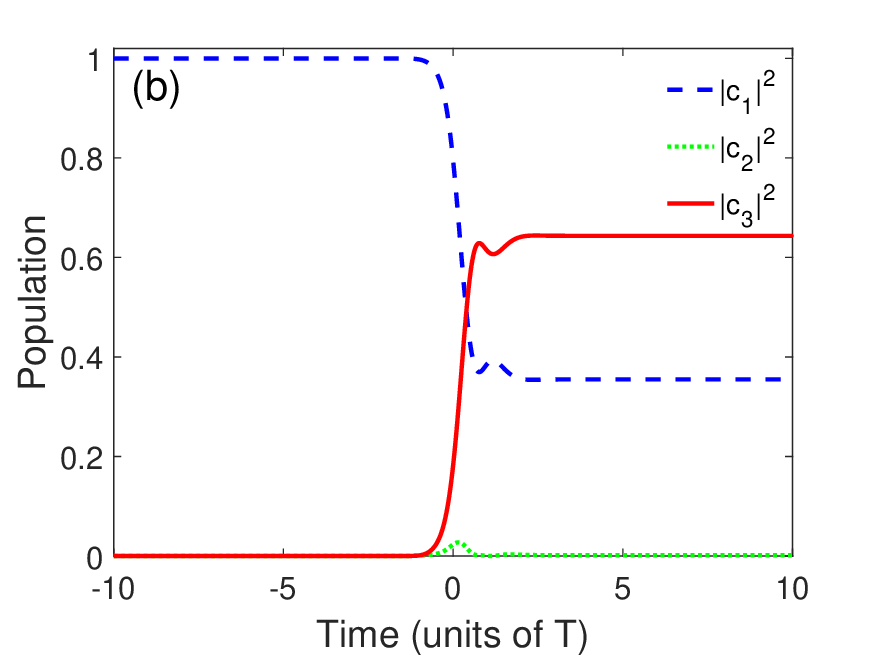}
\caption{Population transfer in the three-level system via the dark-state passage for the f-STA in (a) and for the f-STIRAP in (b). The control parameter is chosen as $\alpha=\arccos(1/\sqrt{3})$, and the other parameters are same with that in Fig.~2(a) and (b).  }
\label{figure3-1}
\end{figure}

The parameter $\alpha$ in the driving pulses is pivotal for the production of the superposition states. The change of the final population on $|3\rangle$ with $\alpha$ is given in Fig.~\ref{figure3}. According to Eq.~(\ref{auxiliarypulse.3}), the population on the state $|3\rangle$ changes with $\alpha$ as the form of $\sin^{2}\alpha$ at the final time, which is confirmed numerically by the red solid line for the f-STA in Fig.~\ref{figure3}. The conversion rate of $0.5$ can be obtained once $\alpha=\pi/4$ or $3\pi/4$ is chosen. By comparison, the change of the population on $|3\rangle$ with $\alpha$ for the f-STIRAP (blue dashed line) deviates from the analytic result. The amplitude and periodic of the oscillation are further reduced. To demonstrate the controllability, the superposition state with unequal population is investigated. As an example, the superposition state with the final population ratio of $\frac{1}{3}:\frac{2}{3}$ between the states $|1\rangle$ and $|3\rangle$ is shown in Fig.~\ref{figure3-1} by choosing the parameter $\alpha=\arccos(1/\sqrt{3})$. It is shown that the population ratio for the f-STA can be controlled precisely, but not for the f-STIRAP. The production of the coherent states by the f-STA is more robust to that by the f-STIRAP under the same parameter condition.

\subsection{C. Effect of the spontaneous emission  }
In the realistic process of preparing the coherent superposition states, the presence of the spontaneous emission in the excited state is inevitable. To obtain the transfer efficiency of the f-STA and the f-STIRAP under the influence of the spontaneous emission, the density matrix based on the quantum master equation is used to describe the change of the population in each energy level of the system. The dynamic evolution of the density operator for the system is described by~\cite{X.Shi2021, P.A.Ivanov2005, M. Scala2011}:
\begin{equation}
\begin{array}{c}i\hbar\dot{\hat{\rho}}=[\widehat{H}(t),\hat{\rho}]+\widehat{D}[\hat{\rho}],
\\
\\ \widehat{D}[\hat{\rho}]=\sum_{i}\gamma_i(2\hat{L}_i\hat{\rho}\hat{L}_i^\dagger-\hat{L}_i^\dagger\hat{L}_i\hat{\rho}-\hat{\rho}\hat{L}_i^\dagger\hat{L}_i).
\end{array} \label{LiouvilleEq}
\end{equation}
$\widehat{D}[\hat{\rho}]$ represents the Lindblad term related to the spontaneous emission in the excited state, which is composed of the spontaneous emission rate $\gamma_i~(i=1,2)$ and the jump operators $\hat{L}_1=\left|1\right\rangle \left\langle2\right|$ and $\hat{L}_2=\left|3\right\rangle\left\langle2\right|$. The spontaneous emission from the state $\left|2\right\rangle$ to the states $\left|1\right\rangle$ and $\left|3\right\rangle$ in the three-level system are denoted by $\gamma_1=i\Gamma_{21}$ and $\gamma_2=i\Gamma_{23}$, respectively. According to Eq.~(\ref{LiouvilleEq}), the Lindblad term for the three-level system is concretely written as:
\begin{equation}
\widehat{D}[\hat{\rho}]=-\frac{i\hbar}{2}\begin{pmatrix}
-2\Gamma_{21}\rho_{22}&(\Gamma_{21}+\Gamma_{23})\rho_{12}&0\\
(\Gamma_{21}+\Gamma_{23})\rho_{21}&2(\Gamma_{21}+\Gamma_{23})\rho_{22}&(\Gamma_{21}+\Gamma_{23})\rho_{23}\\
0&(\Gamma_{21}+\Gamma_{23})\rho_{32}&-2\Gamma_{23}\rho_{22}
\end{pmatrix}.
\end{equation}\label{LindbladD}

\begin{figure}[h]
\centering
\includegraphics[width=0.4\textwidth]{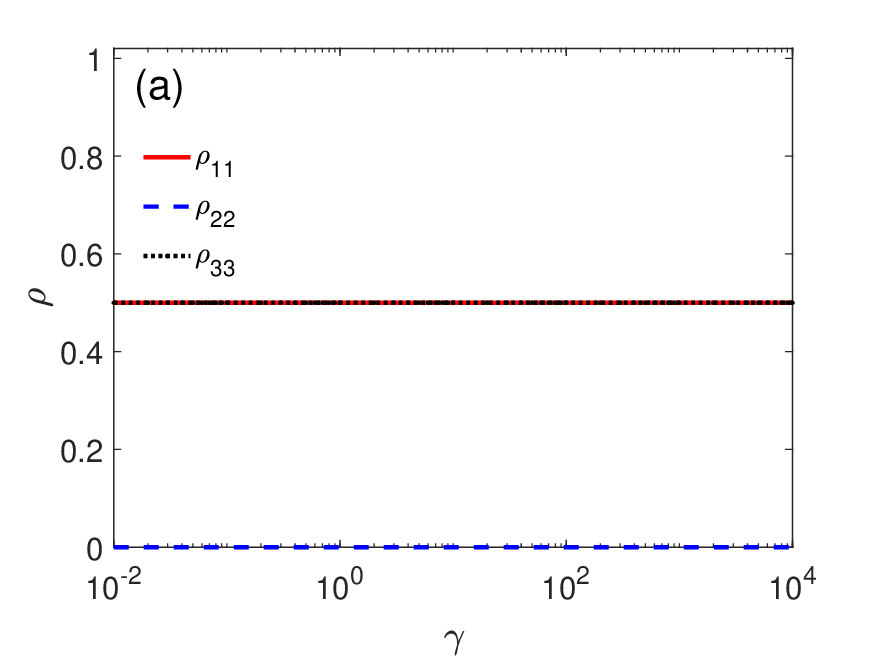}
\includegraphics[width=0.4\textwidth]{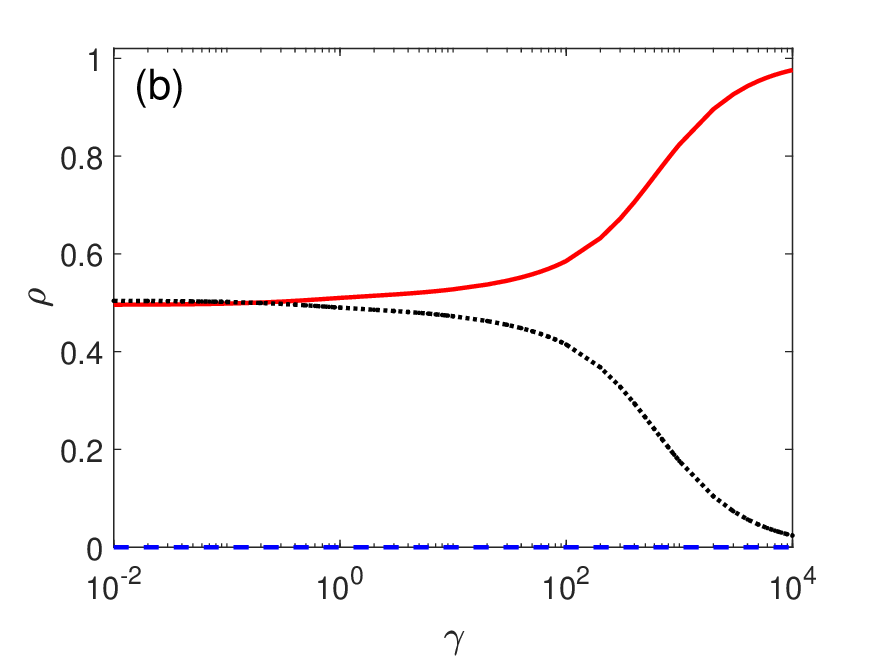}
\caption{Change of the final populations $\rho_{11}$, $\rho_{22}$ and $\rho_{33}$ with the spontaneous emission rate $\gamma$. (a) The case for the f-STA; (b) The case for the f-STIRAP. The other parameters are chosen as $\alpha=\pi/4$, $\Omega_0T=2.0$, $\tau/T=0.7$ and $\Delta=0.2\pi$. }
\label{figure4}
\end{figure}

To simplify the calculation, $\Gamma_{21}=\Gamma_{23}\equiv\Gamma$ is taken, and the dimensionless spontaneous emission rate $\gamma_{1}=\gamma_{2}\equiv\gamma$ is introduced. The change of the components $\rho_{11}$, $\rho_{22}$, and $\rho_{33}$ for the density operator with $\gamma$ by solving Eq.~(\ref{LiouvilleEq}) is demonstrated in Fig.~\ref{figure4}.
Fig.~\ref{figure4}(a) illustrates the case for the f-STA. $\rho_{11}$ and $\rho_{33}$ always keep at $0.5$, which is independent on the spontaneous emission. The reason is that the population on the excited state is always zero, so the spontaneous emission doesn't work for the transfer. Fig.~\ref{figure4}(b) shows the case for the f-STIRAP. $\rho_{11}$ increases and $\rho_{33}$ decreases gradually with increasing $\gamma$ for the f-STIRAP. The population transfer is suppressed by the spontaneous emission rate in this situation. In the process of the dynamical evolution, the spontaneous emission of the population in the excited state $|2\rangle$ to the state $|1\rangle$ is more significant than to the state $|3\rangle$ for larger emission rate. The changes of the population with the spontaneous emission under the counterintuitive pulse sequence for the f-STIRAP and the f-STA are very different. It is evident that the f-STA is more advantageous than the f-STIRAP for preparing coherent superposition with two states in the open systems.

\section*{III. Fractional STA for the four-level system }
\subsection*{A. Theoretical framework }
\begin{figure}[h]
\centering
\includegraphics[width=0.55\textwidth]{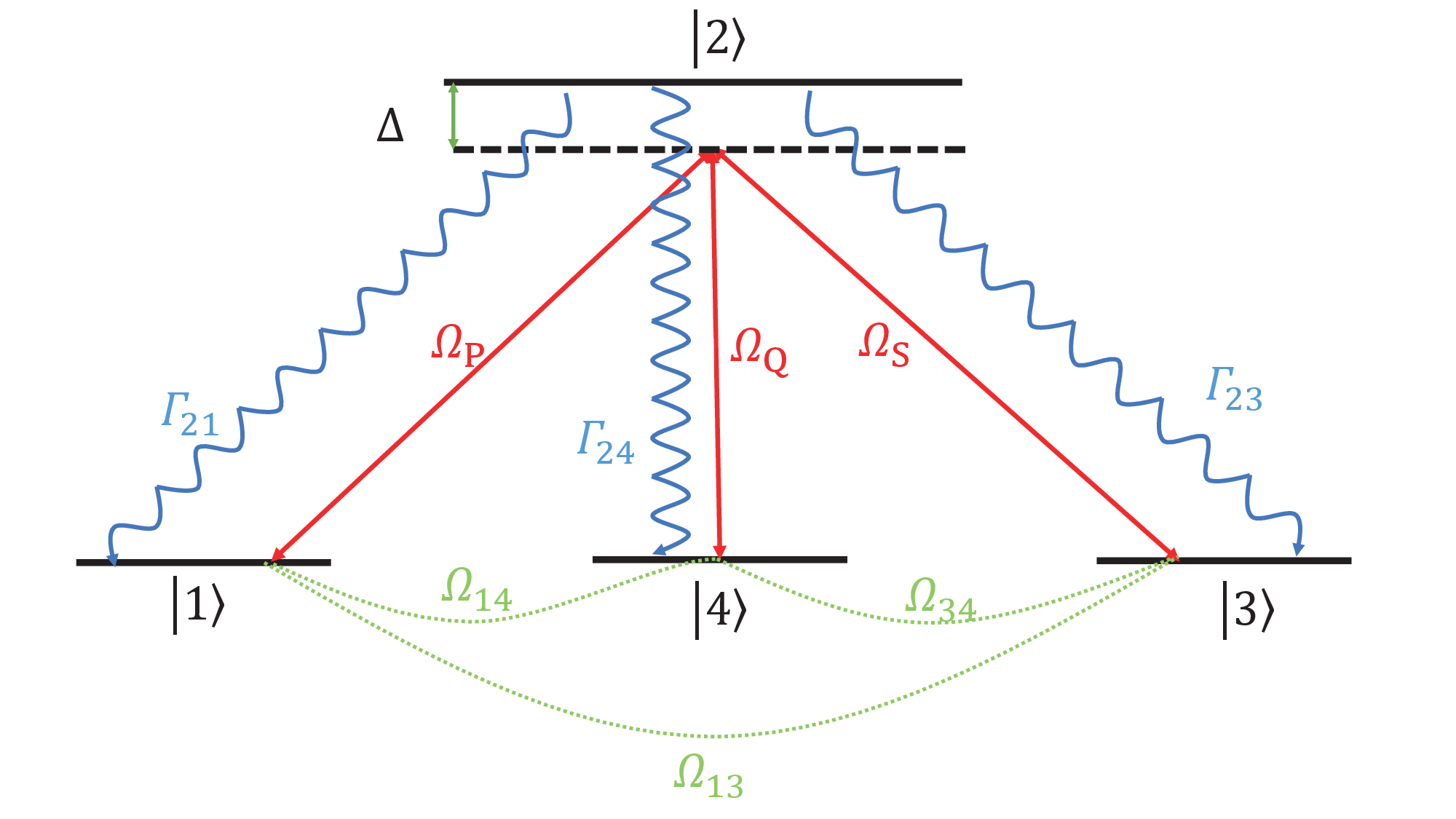}
\caption{Scheme of the population transfer for the four-level system. The coupling scheme is realized by the pump pusle $\Omega_{P}$ and the Stokes pulses $\Omega_{Q}$ and $\Omega_{S}$ (Red solid lines) for the f-STA and the f-STIRAP. The auxiliary pulses $\Omega_{13}$, $\Omega_{14}$ and $\Omega_{34}$ depicted by the dotted green lines are designed only for the f-STA. $\Gamma_{21}$, $\Gamma_{23}$ and $\Gamma_{24}$ represent the spontaneous emissions from the intermediate state $|2\rangle$ to the initial state $|1\rangle$ and the two final states $|3\rangle$ and $|4\rangle$ (Blue wavy lines). }
\label{figure5}
\end{figure}

The scheme for the f-STIRAP and the f-STA in the four-level system with tripod linkage pattern is shown by Fig.~\ref{figure5}. The Hamiltonian in the rotating-wave approximation is written as~\cite{N.V.Vitanov2017, R.Unanyan1998}
\begin{equation}
\widehat{H}_{0}(t)=\frac{\hbar}{2}\left(
\begin{array}{llll}
~~~0&~~~\Omega_{P}(t)&~~~~~~~~~~~0&~~~~~~0\\
\Omega_{P}(t)&~~~~~~\Delta&~~~~~~~~\Omega_{S}(t)&~~~\Omega_{Q}(t)\\
~~~0&~~~\Omega_{S}(t)&~~~~~~~~~~~0&~~~~~~0\\
~~~0&~~~\Omega_{Q}(t)&~~~~~~~~~~~0&~~~~~~0\\
\end{array}
\right),\label{Hamiltonian.4L}
\end{equation}
where $\Omega_{P (S, Q)}(t)$ represents the driving pulses between the pure states $|2\rangle$ and $|1\rangle$ ($|3\rangle$, $|4\rangle$). Similar with the three-level system, here the pure states follow Eq.~(\ref{schrodingereq.1}) but with the probability amplitude $c(t)=\left[c_{1}(t), c_{2}(t), c_{3}(t),c_{4}(t)\right]^{T}$. The system is only populated on the state $|1\rangle$ at the initial time, i.e., $c_{1}(-\infty)=1$, $c_{2}(-\infty)=0$, $c_{3}(-\infty)=0$ and $c_{4}(-\infty)=0$, and the dynamics of the system is investigated by monitoring the populations $|c_{n}(+\infty)|^{2}$ $(n=1, 2, 3,4)$ at $t\rightarrow+\infty$.

The eigenvalues of the Hamiltonian (\ref{Hamiltonian.4L}) are $E_{D1}$=$E_{D2}$=$0$, $E_{B1}=\hbar[\Delta(t)+\sqrt{\Delta^{2}(t)+\Omega^{2}(t)}]/2$ and $E_{B2}=\hbar[\Delta(t)-\sqrt{\Delta^{2}(t)+\Omega^{2}(t)}]/2$ with $\Omega(t)=\sqrt{\Omega_{P}^{2}(t)+\Omega_{S}^{2}(t)+\Omega_{Q}^{2}(t)}$. Accordingly, there are four dressed eigenstates, with two being dark states and two being bright states:
\begin{equation}
\begin{array}{c}
|E_{D1}(t)\rangle=\left(
\begin{array}{l}
~~~~~~\cos\theta(t)\\
~~~~~~~~~~0~~~\\
-\sin\theta(t)\cos\eta(t)\\
-\sin\theta(t)\sin\eta(t)\\
\end{array}
\right),~~~~
|E_{D2}(t)\rangle=\left(
\begin{array}{l}
~~~~0~~~\\
~~~~0~~~\\
\sin\eta(t)\\
-\cos\eta(t)\\
\end{array}
\right),\\
\\
|E_{B1}(t)\rangle=\left(
\begin{array}{l}
~~~~~\sin\phi(t)\sin\theta(t)~~\\
~~~~~~~~~\cos\phi(t)~~~\\
\sin\phi(t)\cos\theta(t)\cos\eta(t)\\
\sin\phi(t)\cos\theta(t)\sin\eta(t)\\
\end{array}
\right), ~~~~|E_{B2}(t)\rangle=\left(
\begin{array}{l}
~~~~~\cos\phi(t)\sin\theta(t)~~\\
~~~~~~~~~-\sin\phi(t)~~~\\
\cos\phi(t)\cos\theta(t)\cos\eta(t)\\
\cos\phi(t)\cos\theta(t)\sin\eta(t)\\
\end{array}
\right).\\
\end{array}\label{DBStates.4L}
\end{equation}
The mixing angles are determined by $\tan\eta(t)=\Omega_{Q}(t)/\Omega_{S}(t)$, $\tan\theta(t)=\Omega_{P}(t)/\sqrt{\Omega_{S}^{2}(t)+\Omega_{Q}^{2}(t)}$ and $\phi=\arctan(\sqrt{\Omega_{P}^{2}(t)+\Omega_{S}^{2}(t)+\Omega_{Q}^{2}(t)}/\Delta)/2$. When the system is in the dark state $|E_{D1}(t)\rangle$, the superposition of $\left | 1 \right \rangle$, $\left | 3 \right \rangle$, and $\left | 4 \right \rangle$ with arbitrary proportion can be determined by the mixing angles $\theta$ and $\eta$. The proportion will be $\cos^{2}\theta$ in the state $\left | 1 \right \rangle$, $\sin^{2}\theta\cos^{2}\eta$ in the state $\left|3\right\rangle$, and $\sin^{2}\theta\sin^{2}\eta$ in the state $\left|4\right\rangle$ for the whole process. The specific form of the driving pulses for the four-level system are chosen as
\begin{equation}\nonumber
\Omega_{P}(t)=\Omega_{0}\sin\beta e^{-(t-\tau)^{2}/T^{2}},
\end{equation}
\begin{equation}
\Omega_{S}(t)=\Omega_{0}[e^{-(t+\tau)^{2}/T^{2}}+\cos\beta\cos\chi e^{-(t-\tau)^{2}/T^{2}}],
\label{fourL-PulseS}
\end{equation}
\begin{equation}\nonumber
\Omega_{Q}(t)=\Omega_{0}[e^{-(t+\tau)^{2}/T^{2}}+\cos\beta\sin\chi e^{-(t-\tau)^{2}/T^{2}}].
\end{equation}
The final proportion of the superposition state is determined by the controllable parameters $\beta$ and $\chi$ in the pulses.
Associated with the definition of the mixing angles, we have $\theta(t\rightarrow+\infty)=\beta$ and $\eta(t\rightarrow+\infty)=\chi$.

Now, we will demonstrate the process for obtaining the auxiliary pulses for implementing the f-STA in the four-level system. Based on Eq.~(\ref{DBStates.4L}), the unitary transformation matrix of the four-level system is constructed as
\begin{equation}
\widehat{U}(t)=\left(
\begin{array}{llll}
~~~~~~~\cos\theta(t)&~~~~~~~0&~~~~~~~~~~~~~~~~~\sin\phi(t)\sin\theta(t)&~~~~~~~~~~~\cos\phi(t)\sin\theta(t)\\
~~~~~~~~~~~0&~~~~~~~0&~~~~~~~~~~~~~~~~~~~~~\cos\phi(t)&~~~~~~~~~~~~~~-\sin\phi(t)\\
~~~-\sin\theta(t)\cos\eta(t)&~~~\sin\eta(t)&~~~~~~~~~~~\sin\phi(t)\cos\theta(t)\cos\eta(t)&~~~~~~\cos\phi(t)\cos\theta(t)\cos\eta(t)\\
~~~-\sin\theta(t)\sin\eta(t)&~-\cos\eta(t)&~~~~~~~~~~~\sin\phi(t)\cos\theta(t)\sin\eta(t)&~~~~~~\cos\phi(t)\cos\theta(t)\sin\eta(t)\\
\end{array}
\right).\label{UMatrix4}
\end{equation}
The auxiliary Hamiltonian according to Eq.~(\ref{auxiliarypulse}) is
\begin{equation}
\widehat{H}_{a}=i\left(\begin{array}{cccc}0 & \Omega_{12}(t) & \Omega_{13}(t) & \Omega_{14}(t)
 \\ -\Omega_{12}(t) & 0 & \Omega_{23}(t) & \Omega_{24}(t)
 \\ -\Omega_{13}(t) & -\Omega_{23}(t)  & 0 & \Omega_{34}(t)
 \\ -\Omega_{14}(t) & -\Omega_{24}(t) & -\Omega_{34}(t) & 0\end{array}\right),
\label{auxiliaryHamiltonian.4L}
\end{equation}
and the specific forms of the auxiliary pulses are
\begin{equation}
\begin{array}{c}
\Omega_{12}(t)=\dot{\phi}(t)\sin\theta(t),~\Omega_{13}(t)=\dot{\theta}(t) \cos\eta(t), ~\Omega_{14}(t)=\dot{\theta}(t) \sin \eta(t),
\\
\\
\Omega_{23}(t)= -\dot{\phi}(t)\cos\theta(t)\cos\eta(t),
~\Omega_{24}(t)= -\dot{\phi}(t)\cos\theta(t)\sin\eta(t),
~\Omega_{34}(t)=-\dot{\eta}(t),
\end{array}\label{auxiliarypulse.5L}
\end{equation}
with $\dot{\theta}=[\dot{\Omega}_{P}(\Omega_{S}^{2}+\Omega_{Q}^{2})-\Omega_{p}(\Omega_{S} \dot{\Omega}_{S}+\Omega_{Q} \dot{\Omega}_{Q})]/[(\Omega_{S}^{2}+\Omega_{Q}^{2}+\Omega_{P}^{2}) \sqrt{\Omega_{S}^{2}+\Omega_{Q}^{2}}]$, $\dot{\eta}=(\dot{\Omega}_{Q} \Omega_{S}-\dot{\Omega}_{S} \Omega_{Q})/(\Omega_{S}^{2}+\Omega_{Q}^{2})$, and $\dot{\phi}=\Delta(\dot{\Omega}_{P} \Omega_{P}+\dot{\Omega}_{S}\Omega_{S}+\dot{\Omega}_{Q}\Omega_{Q})/[\Omega(\Delta^{2}+4 \Omega^{2})]$.
Because $|E_{D1}(t)\rangle$ and $|E_{D2}(t)\rangle$ are degenerate, the nonadiabatic coupling between them can't be suppressed, even in the adiabatic limit. The nonadiabatic coupling between $|E_{D1}(t)\rangle$ and $|E_{D2}(t)\rangle$
causes a transition between them with probability $\sin^{2}[\int_{-\infty}^{+\infty}\dot{\eta}(t)\sin\theta(t)dt]$~\cite{R.Unanyan1998}.
When the mixing angle $\eta(t)$ is chosen time-independent, there is no any transition between them. The population transfer of the system will only keep in the dark-state passage given by $|E_{D1}(t)\rangle$. The superposition of the states $\left | 1 \right \rangle$, $\left | 3 \right \rangle$ and $\left | 4 \right \rangle$ with arbitrary proportion can be realized by choosing suitable $\theta$ and $\eta$.

\subsection{B. Coherent production of the superposition state}
We first demonstrate the case of the superposition of the states $\left|1\right\rangle$, $\left|3\right\rangle$ and $\left|4\right\rangle$ with equal proportion. The final state is $\left|\psi(+\infty)\right\rangle=(\left|1\right\rangle+\left| 3\right\rangle+\left| 4\right\rangle)/\sqrt{3}$ and the ratio of the populations on $\left|1\right\rangle$, $\left|3\right\rangle$ and $\left|4\right\rangle$ is $\frac{1}{3}:\frac{1}{3}:\frac{1}{3}$. The controllable parameters $\beta=\arccos(1/\sqrt{3})$ and $\chi=\pi/4$ are selected for realizing the equal superposition of the three low-energy states. This superposition state is useful for building three-bit quantum gates for quantum computation~\cite{BTTorosov2020,SGStanchev2024}.

\begin{figure}[h]
\centering
\includegraphics[width=0.35\textwidth]{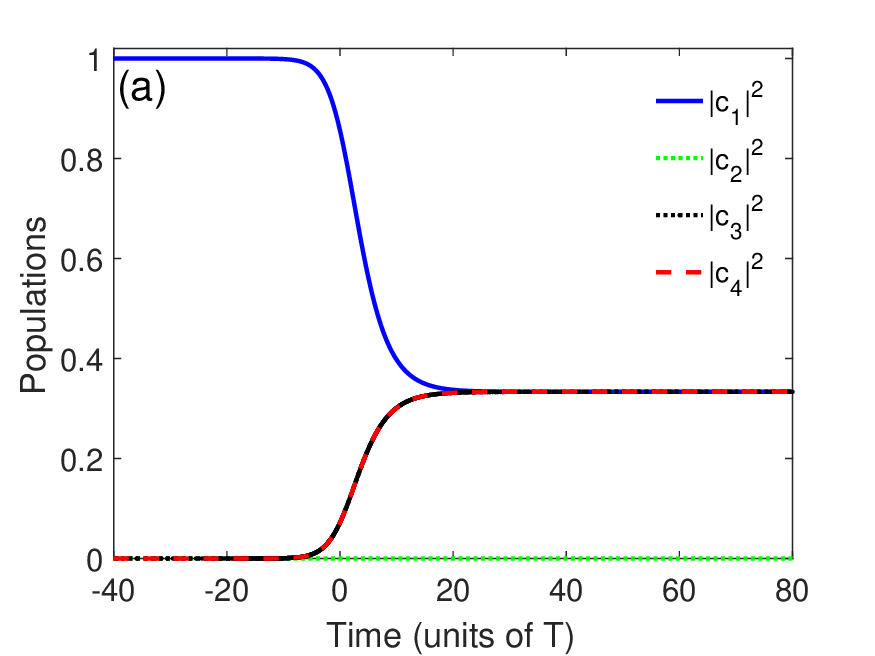}
\includegraphics[width=0.35\textwidth]{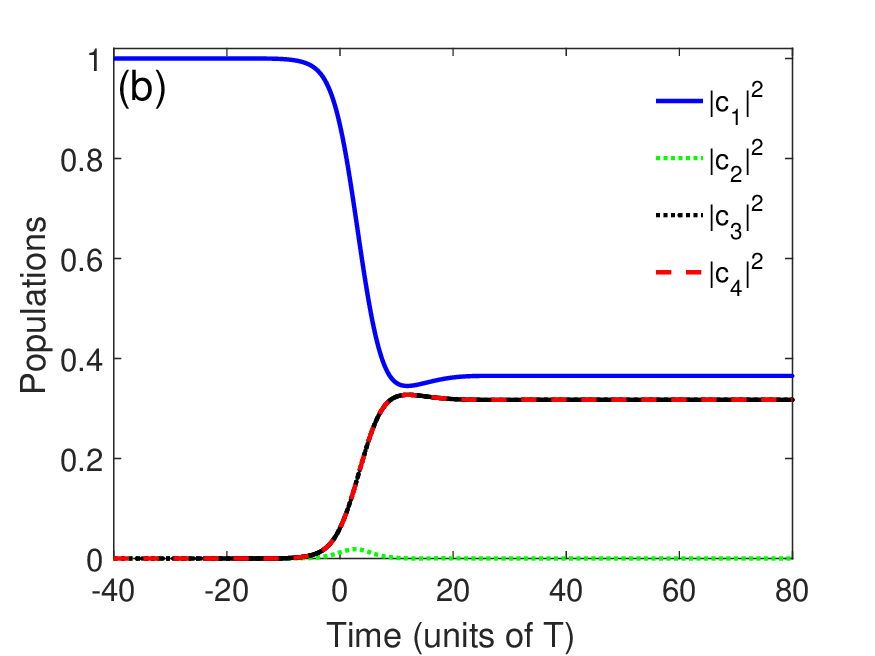}
\includegraphics[width=0.35\textwidth]{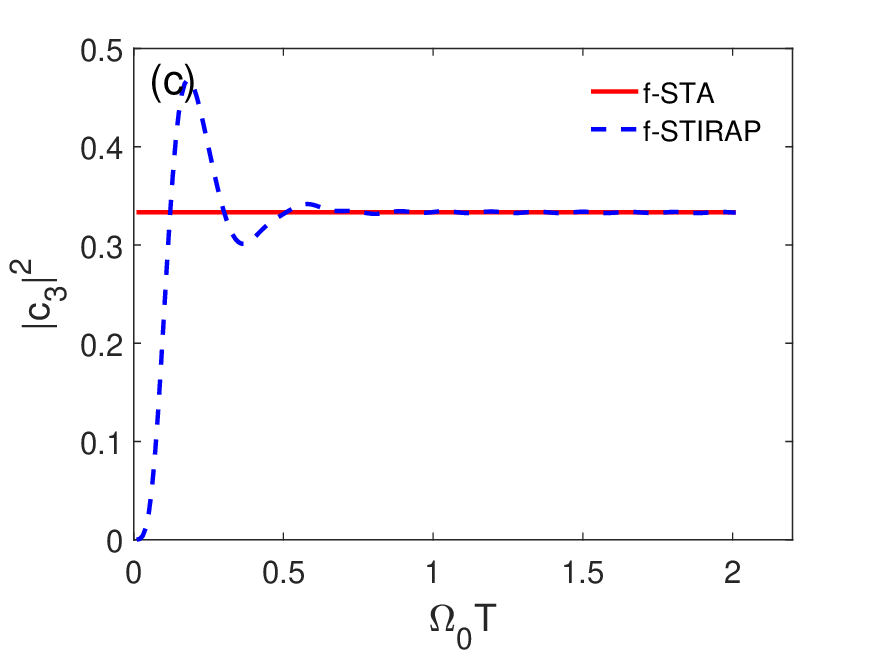}
\includegraphics[width=0.35\textwidth]{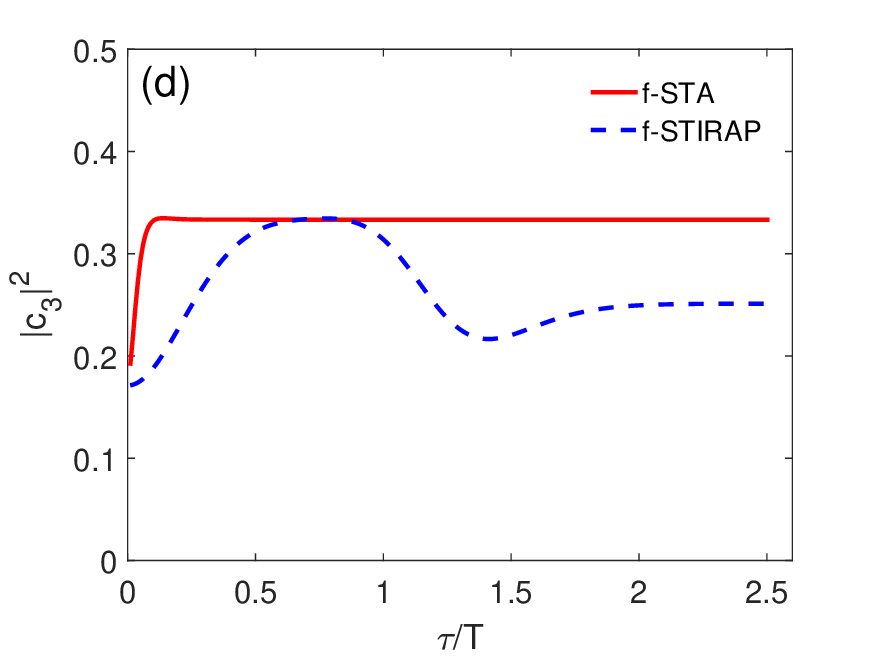}
\caption{Population transfer in the four-level system by the dark-state passage for the f-STA in (a) and for the f-STIRAP in (b). The parameters are chosen as $\beta=\arccos(1/\sqrt{3})$, $\chi=\pi/4$, $\Omega_0T=2.0$, $\tau/T=0.7$ and $\Delta=0.2\pi$. The effects of the peak pulse intensity $\Omega_{0}$ in (c) at $\tau/T=0.7$ and the time delay $\tau$ in (d) at $\Omega_{0}T=2.0$ on the population of the state $|3\rangle$ for both of the f-STA (Red solid line) and the f-STIRAP (Blue dashed line).  }
\label{figure6}
\end{figure}

The evolution of the populations in the four-level system via the dark-state passage $|E_{D1}(t)\rangle$ is demonstrated in Fig.~\ref{figure6}(a) and (b). Fig.~\ref{figure6}(a) illustrates the population transfer for the f-STA from the initial state $|1\rangle$ to the two final ones $|3\rangle$ and $|4\rangle$. The population is on the state $|1\rangle$ at the initial time, and then $1/3$ of it is transferred to the state $|3\rangle$ and $1/3$ to $|4\rangle$ at the final time. There is no population on $|2\rangle$ during the whole process. The final populations on $|3\rangle$ and $|4\rangle$ don't reach to $1/3$ for the f-STIRAP in Fig.~\ref{figure6}(b). The f-STA introduces a group of auxiliary pulses (\ref{auxiliarypulse.5L}) to eliminate the nonadiabatic transitions, perfectly achieving a coherent superposition state of $|1\rangle$,  $|3\rangle$ and $|4\rangle$ with equal proportion.

The effects of the peak pulse intensity and the time delay on the final populations is illustrated in Fig.~\ref{figure6}(c) and (d). Fig.~\ref{figure6}(c) shows the change of the final population on $\left|3\right\rangle$ with the peak pulse intensity $\Omega_{0}$. The final population on $\left|3\right\rangle$ with $\Omega_{0}$ for the f-STIRAP exhibits oscillation around $1/3$. The amplitude of such oscillation decreases with further increasing the pulse intensity. By comparison, due to bypassing the adiabatic requirement by the auxiliary pulses, the f-STA consistently achieves a transfer efficiency of $1/3$ regardless of the magnitude of the pulse intensity. Fig.~\ref{figure6}(d) shows the change of the final population on $\left|3\right\rangle$ with the time delay $\tau$. The final population on $\left|3\right\rangle$ is increased to $1/3$ for the f-STA when the time delay is larger than a small critical value. For even larger time delay, the population keeps at $1/3$ and is independent on $\tau$. The population transfer can accomplish even if the driving pulses don't overlap due to the presence of the auxiliary pulses. The population on $\left|3\right\rangle$ for the f-STIRAP approaches to about $1/3$ when $\tau$ is at a small range near $0.7T$. The population deviates largely from $1/3$ for the other part of the time delay. The population on $\left|4\right\rangle$ changes synchronously with that on $\left|3\right\rangle$ for the chosen parameter $\chi=\pi/4$, so the corresponding results haven't been shown.

\begin{figure}[h]
\centering
\includegraphics[width=0.8\textwidth]{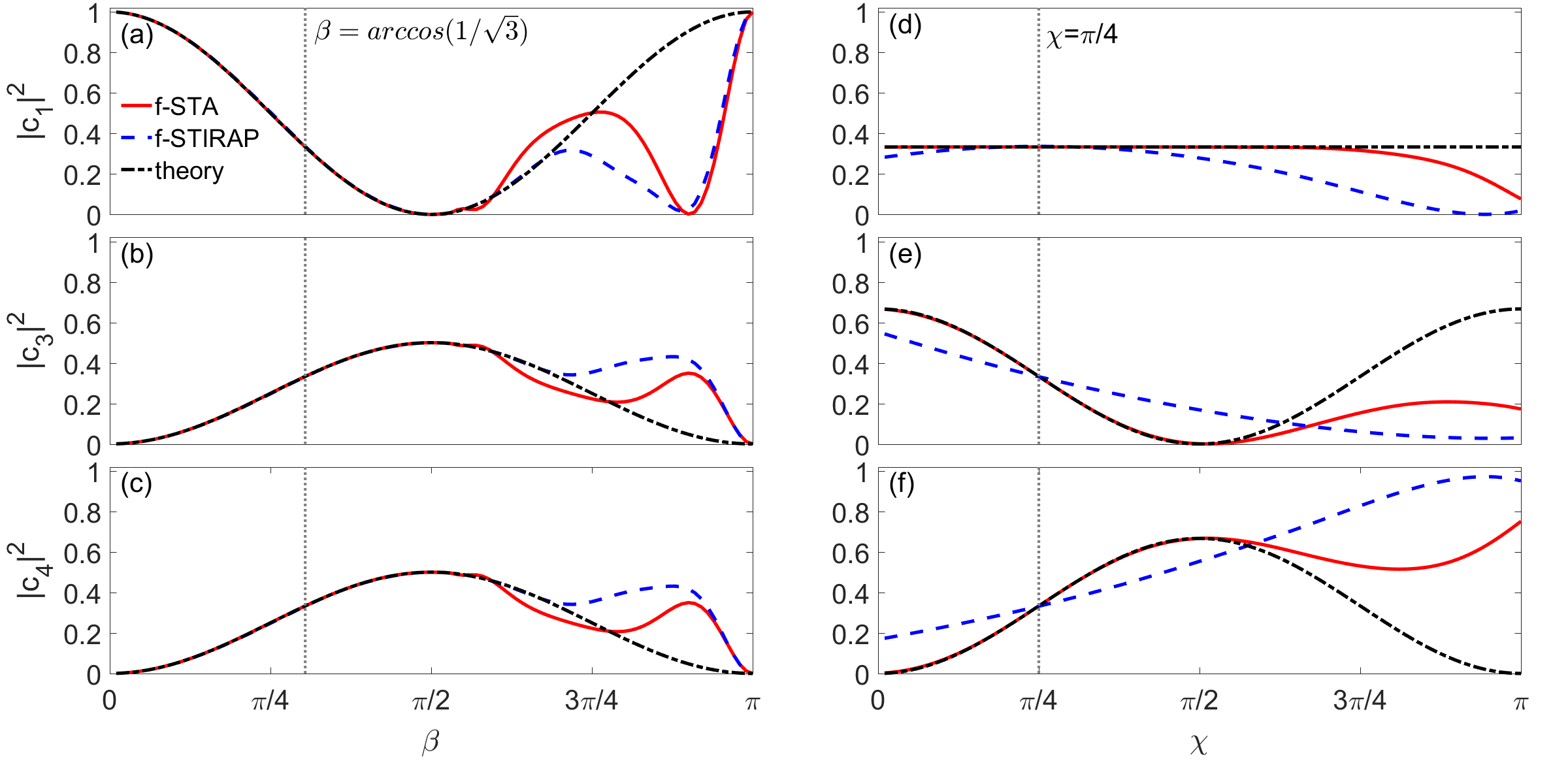}
\caption{Change of the final populations on $|1\rangle$, $|3\rangle$ and $|4\rangle$ with $\beta$ in (a-c) and $\chi$ in (d-f) for the f-STA (Red solid line) and the f-STIRAP (Blue dashed line). The dotted-dashed lines give the corresponding theoretical results. The vertical dotted line corresponds to $\beta=\arccos(1/\sqrt{3})$ in (a-c) and $\chi=\pi/4$ in (d-f), respectively. The other parameters are chosen as $\Omega_0T=2.0$, $\tau/T=0.7$ and $\Delta=0.2\pi$. }
\label{figure7}
\end{figure}

\begin{figure}[h]
\centering
\includegraphics[width=0.4\textwidth]{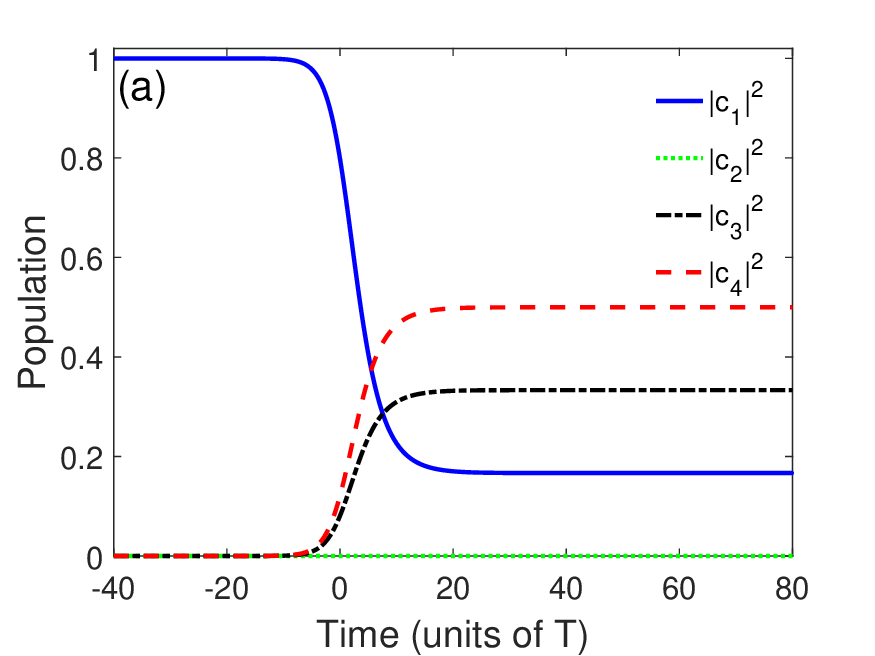}
\includegraphics[width=0.4\textwidth]{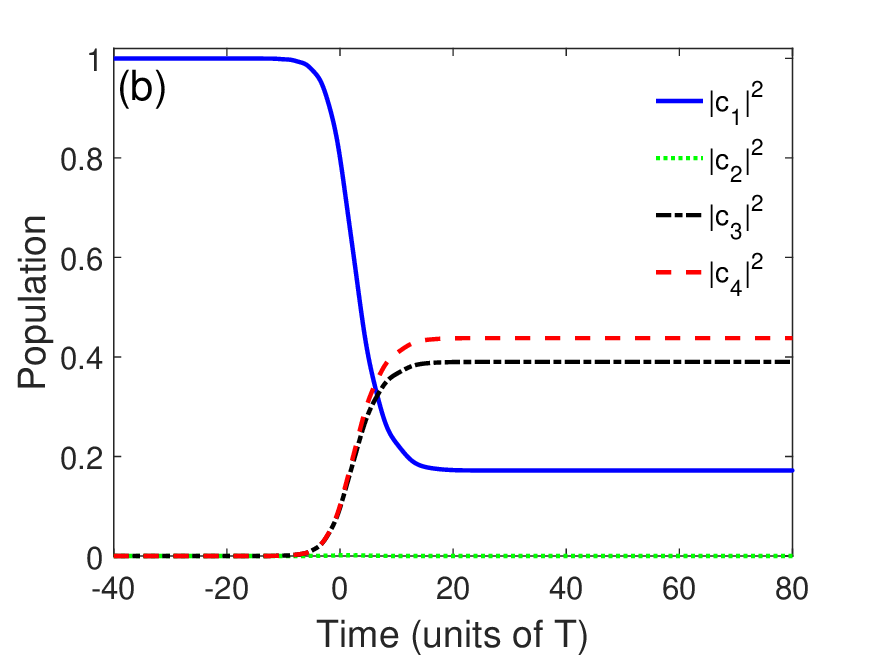}
\caption{Population transfer in the four-level system by the dark-state passage for the f-STA in (a) and for the f-STIRAP in (b). The control parameters are chosen as $\beta=\arccos(1/\sqrt{6})$ and $\chi=\arccos(\sqrt{2/5})$. The other parameters are same with that in Fig.~7(a) and (b).}
\label{figure7-1}
\end{figure}

Unlike the three-level system, there're two controllable parameters $\beta$ and $\chi$ in the driving pulses for the four-level system in order to produce the superposition states. The changes of the final populations on $\left|1\right\rangle$, $\left| 3\right\rangle$ and $\left| 4\right\rangle$ with $\beta$ are demonstrated in Fig.~\ref{figure7}(a), (b) and (c). For $\chi=\pi/4$, the theoretical populations are $\cos^{2}\beta$ on $\left|1\right\rangle$ and $\sin^{2}\beta/2$ on $\left|3\right\rangle$ and $\left|4\right\rangle$ since $\theta(t\rightarrow+\infty)=\beta$ is taken. It is found that the final populations for the f-STA and the f-STIRAP are consistent with the theoretical values when $\beta$ is within the range of $0$ to $\pi/2$. When $\beta$ is at the range of $\pi/2$ to $\pi$, the results for both the f-STA and the f-STIRAP deviate from the theoretical values. The change of the final populations on $\left|1\right\rangle$, $\left| 3\right\rangle$ and $\left| 4\right\rangle$ with $\chi$ is demonstrated in Fig.~\ref{figure7}(d), (e) and (f). For $\beta=\arccos(1/\sqrt{3})$, the theoretical populations are $1/3$ on $\left|1\right\rangle$, $(2\cos^{2}\chi)/3$ on $\left|3\right\rangle$ and $(2\sin^{2}\chi)/3$ on $\left|4\right\rangle$. When the parameter $\chi$ is chosen at the range of $0$ to $\pi/2$, the change of the populations with $\chi$ for the f-STA agrees with the theoretical value, while the population for the f-STIRAP doesn't align with the theoretical value. When the parameter $\chi$ is at the range of $\pi/2$ to $\pi$, the populations for the f-STA and the f-STIRAP no longer fully agree with the theoretical value. When the two parameters are selected in the range of $0$ to $\pi/2$, the f-STA can achieve any proportional superposition of $\left| 1\right\rangle$, $\left| 3\right\rangle$ and $\left| 4\right\rangle$, but the f-STIRAP can't do that. As an example, the superposition state with the final population ratio of $\frac{1}{6}:\frac{1}{3}:\frac{1}{2}$ between the states $|1\rangle$, $|3\rangle$ and $|4\rangle$ is shown in Fig.~\ref{figure7-1} by choosing the parameters $\beta=\arccos(1/\sqrt{6})$ and $\chi=\arccos(\sqrt{2/5})$. This fully demonstrates the stability and controllability of the f-STA in the realization of the superposition states with three components.

\subsection{C. Effect of the spontaneous emission}
When the spontaneous emission is considered, the population transfer can be investigated by solving the quantum master equation~(\ref{LiouvilleEq}) of the density operators. For the four-level system, the spontaneous emissions from the state $\left|2\right\rangle$ to the other states are denoted as $\gamma _{i}$ $(i=1,2,3)$, where $\gamma _{1}=i\Gamma _{21}$, $\gamma _{2}=i\Gamma _{23}$ and $\gamma _{3}=i\Gamma _{24}$. The jump operators are defined as $\hat{L}_{1}=\left | 1 \right \rangle \left \langle 2 \right |$, $\hat{L}_{2}=\left | 3 \right \rangle \left \langle 2 \right |$ and $\hat{L}_{3}=\left | 4 \right \rangle \left \langle 2 \right |$. According to Eq.~(\ref{LiouvilleEq}), the following Lindblad term can be expressed as:
\begin{equation}
\begin{array}{c}
\\ \widehat{D}[\hat{\rho}]=-\frac{i\hbar}{2}\left(
\begin{array}{cccc}-2
\Gamma_{21} \rho_{22} & \left(\Gamma_{21}+\Gamma_{23}+\Gamma_{24}\right) \rho_{12} & 0 & 0 \\ \left(\Gamma_{21}+\Gamma_{23}+\Gamma_{24}\right) \rho_{21} & 2\left(\Gamma_{21}+\Gamma_{23}+\Gamma_{24}\right) \rho_{22} & \left(\Gamma_{21}+\Gamma_{23}+\Gamma_{24}\right) \rho_{23} & \left(\Gamma_{21}+\Gamma_{23}+\Gamma_{24}\right) \rho_{24} \\
0 & \left(\Gamma_{21}+\Gamma_{23}+\Gamma_{24}\right) \rho_{32} & -2 \Gamma_{23} \rho_{22} & 0 \\
0 & \left(\Gamma_{21}+\Gamma_{23}+\Gamma_{24}\right) \rho_{42} & 0 & -2 \Gamma_{24} \rho_{22}
\end{array}\right)
\end{array}, \label{Lindblad.4L}
\end{equation}
It is supposed that $\Gamma=\Gamma_{21}=\Gamma_{23}=\Gamma_{24}$ is satisfied and $\gamma=\gamma_{1}=\gamma_{2}=\gamma_{3}$ represents the dimensionless spontaneous emission rate.

\begin{figure}[h]
\centering
\includegraphics[width=0.4\textwidth]{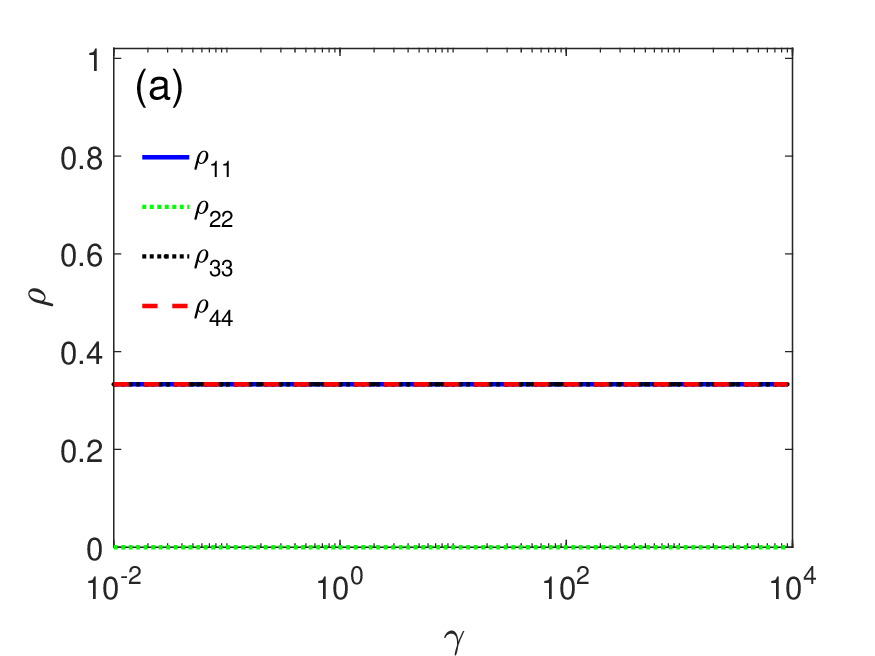}
\includegraphics[width=0.4\textwidth]{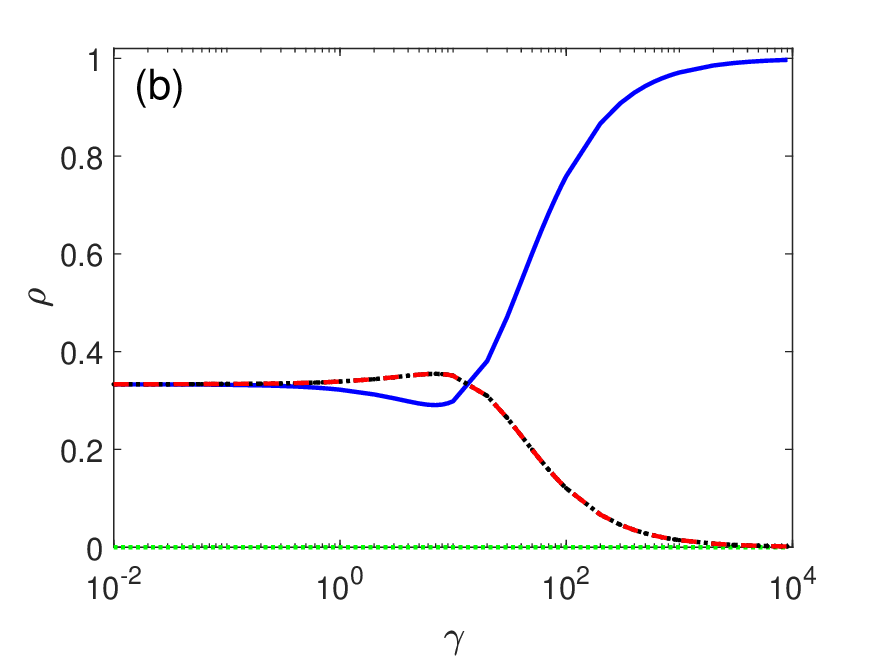}
\caption{Change of the final populations $\rho_{11}$, $\rho_{22}$, $\rho_{33}$ and $\rho_{44}$ with the spontaneous emission rate $\gamma$: (a) The case for the f-STA; (b) The case for the f-STIRAP. The other parameters are chosen as $\beta=\arccos(1/\sqrt{3})$, $\chi=\pi/4$, $\Omega_0T=2.0$, $\tau/T=0.7$ and $\Delta=0.2\pi$.  }
\label{figure8}
\end{figure}

The effect of the spontaneous emission on the population transfer for the four-level system are shown in Fig.~\ref{figure8}.
Fig.~\ref{figure8}(a) illustrates the change of the populations with the spontaneous emission rate $\gamma$ for the f-STA. The components of the density operator $\rho_{11}$, $\rho_{33}$ and $\rho_{44}$ always keep at $1/3$, which is independent on the spontaneous emission. There's no population on $\rho_{22}$ during the whole transfer process. Fig.~\ref{figure8}(b) illustrates the change of the populations with $\gamma$ for the f-STIRAP. When the spontaneous emission rate is sufficiently small, the population on the components $\rho_{11}$, $\rho_{33}$ and $\rho_{44}$ almost is $1/3$. The mechanism of the f-STIRAP plays a key role. For even larger $\gamma$ (overdamping), the population is dominated by the spontaneous emission. The transfer process is suppressed and the population mainly stays on the sate $\left| 1\right\rangle$. For the intermediate value of the spontaneous emission rate ($\gamma\sim1$), there exists some competition between the f-STIRAP and the spontaneous emission. The f-STA is more advantageous than the f-STIRAP for preparing coherent superposition with three states in the open systems.

\subsection{Conclusions }
The scheme for realizing the coherent superposition states in a three-level system and a four-level system by the f-STA is proposed definitely. For the three-level system with $\Lambda$-type linkage pattern, the superposition state can be implemented by an auxiliary Hamiltonian, which is determined by the two driving pulses with a specific value of the controllable parameter $\alpha$. For the four-level system with tripod structure, the superposition states can be realized by another auxiliary Hamiltonian, which can be derived by designing three suitable driving pulses with specific values of the two controllable parameters $\beta$ and $\chi$. The effects of the peak pulse intensity and the time delay of the driving pulses on the population transfer are studied by solving the time-dependent Schr$\ddot{o}$dinger equations. Compared to the f-STIRAP, the transfer process for the f-STA is not limited by the requirement for the adiabaticity of the system. The effectiveness of utilizing the f-STA to prepare two- and three-state superpositions in the three- and four-level systems is investigated, taking into account the influence of the spontaneous emission from the intermediate state. The spontaneous emission has a suppressive effect on the preparation of the superposition states in the f-STIRAP, but it has little impact on the f-STA. The f-STA allows for the realization of arbitrary proportion of superposition states through controlling of the parameters of the driving pulses. The fast and robust production of the superposition states by the f-STA is verified and the mechanism can be generalized to produce the superposition states with more than three components in the future.

\section*{Acknowledgments}
This work is supported by the NSF of China (Grant No.~11405100), the Natural Science Basic Research Program in Shaanxi Province of China (Grant Nos.~2020JM-507 and~2019JM-332).

\end{document}